\documentclass[aps, twocolumn, pra, amssymb, amsmath, superscriptaddress,preprintnumbers]{revtex4-2}
\usepackage{graphicx}
\usepackage{dcolumn}
\usepackage{bm}
\usepackage{amsmath}
\usepackage{subfigure}
\usepackage{amsfonts}
\usepackage{xcolor}
\usepackage{appendix}
\usepackage{multirow}
\usepackage{booktabs}
\usepackage{tabularx}
\usepackage[colorlinks,
            linkcolor=blue,
            anchorcolor=red,
            citecolor=blue
            ]{hyperref}
\usepackage{amsthm}
\usepackage{setspace}
\usepackage{lineno}

\begin{document}
\title{Diverse methods and practical aspects in controlling single semiconductor qubits: a review}
\author{Jia-Ao Peng}\thanks{These authors contributed equally to this work}
\affiliation{State Key Laboratory of Superlattices and Microstructures, Institute of Semiconductors, Chinese Academy of Sciences, Beijing 100083, China}
\affiliation{College of Materials Science and Opto-Electronic Technology, University of Chinese Academy of Sciences, Beijing 100049, China}

\author{Chu-Dan Qiu}\thanks{These authors contributed equally to this work}
\affiliation{State Key Laboratory of Superlattices and Microstructures, Institute of Semiconductors, Chinese Academy of Sciences, Beijing 100083, China}
\affiliation{College of Materials Science and Opto-Electronic Technology, University of Chinese Academy of Sciences, Beijing 100049, China}

\author{Wen-Long Ma}
\email{wenlongma@semi.ac.cn}
\affiliation{State Key Laboratory of Superlattices and Microstructures, Institute of Semiconductors, Chinese Academy of Sciences, Beijing 100083, China}
\affiliation{College of Materials Science and Opto-Electronic Technology, University of Chinese Academy of Sciences, Beijing 100049, China}

\author{Jun-Wei Luo}
\email{jwluo@semi.ac.cn}
\affiliation{State Key Laboratory of Superlattices and Microstructures, Institute of Semiconductors, Chinese Academy of Sciences, Beijing 100083, China}
\affiliation{College of Materials Science and Opto-Electronic Technology, University of Chinese Academy of Sciences, Beijing 100049, China}

\begin{abstract}
Quantum control allows a wide range of quantum operations employed in molecular physics, nuclear magnetic resonance and quantum information processing. Thanks to the existing microelectronics industry, semiconducting qubits, where quantum information is encoded in spin or charge degree freedom of electrons or nuclei in semiconductor quantum dots, constitute a highly competitive candidate for scalable solid-state quantum technologies. In quantum information processing, advanced control techniques are needed to realize quantum manipulations with both high precision and noise resilience. In this review, we first introduce the basics of various widely-used control methods, including resonant excitation, adabatic passage, shortcuts to adiabaticity, composite pulses, and quantum optimal control. Then we review the practical aspects in applying these methods to realize accurate and robust quantum gates for single semiconductor qubits, such as Loss-DiVincenzo spin qubit, spinglet-triplet qubit, exchange-only qubit and charge qubit. 

\end{abstract}
\maketitle

\section{INTRODUCTION}
Quantum control aims to drive the state of a quantum system towards a target one or realize a specific quantum operation mostly via an external engineered electromagnetic field. It has been a long history to strive for ideal quantum control against hardware imperfections and external noise \cite{Rabitz2000,Peirce1988}. A variety of efficient and robust quantum control methods have been proposed and widely employed in molecular physics \cite{MS2003,Koch2019}, nuclear magnetic resonance (NMR) \cite{Ota2009,Lapert2012} and quantum information processing \cite{Nielson2000,Damme2017,Iva2015,Cohen2016,Morton2005}.

There are several widely-used quantum control methods. For a simple two-level quantum system, the resonant excitation (RE) method is the simplest one and can be fast if the pulses are intense enough. Extending from RE, the composite pulses (CPs) method, which consists of a sequence of resonant pulses, was developed in the early eighties \cite{Lev1982,Freeman1980}. There are renewed interests on the CPs method due to its generality and universality, as demonstrated in \cite{Tor2018,Dou2016}. The adiabatic passage method, including rapid adiabatic passage (RAP), stimulated Raman adiabatic passage (STIRAP) and their variants, can produce effective and robust control but require long time for the adiabatic transfer \cite{Ig2018,Ds2020}. To pursue both robustness and speed, another important class of methods called shortcuts to adiabaticity (STA) has been extensively studied in the last decade \cite{Ds2020,DG2019}. In addition, the need for more precise control over quantum dynamics has led to the development of advanced control methods based on optimal control theory. Although only numerical solutions are given, quantum optimal control (QOC) can provide an optimal control field modulation that most precisely implements a desired quantum operation, with minimum energy consumption and maximum robustness against hardware imperfections and external noises \cite{Mah2023}.

Semiconductor qubits are versatile for a variety of quantum applications, particularly in the areas of quantum simulation \cite{Barthelemy2013}, sensing \cite{Chern2019}, computation and communication \cite{Scappucci2021}. In particular, due to the compatibility with current semiconductor microelectronics technologies, semiconductor-based quantum information processing is highly promising to realize a large-scale fault-tolerant quantum computer. However, almost all of these applications require accurate and robust quantum control of single semiconductor qubits. In this review, we will review the principles and applications of various control methods to single semiconductor qubits. After providing an overview of five widely-used quantum control methods, we elaborate on the application of these methods in controlling single semiconductor qubits. 

The physical platform we focus on is semiconductor quantum dot (QD) systems. Specifically, there are various types of qubits to encode quantum information in the spin or charge states of electrons or holes in QDs, including Loss-DiVincenzo spin qubit \cite{Lo1998}, singlet-triplet qubits \cite{Le2002}, triple-dot exchange-only qubits \cite{Ba2000,Ni2000}, charge qubits \cite{Ha2003} and hybrid qubits \cite{Shi2012}. Coherent spin manipulation in QDs can be realized via via magnetic, electric, or spin-orbit coupling control, where the latter two approaches are more promising due to the fast speed and potential scalability \cite{RH2007,HB2010,BM2012}.

There are already many excellent reviews about semiconductor-based quantum information processing \cite{kloeffel2013,Chatterjee2021,Burkard2023,Michielis2023}, while this review aims to give a self-contained introduction to various single-qubit quantum control methods, and discuss their practical applications in semiconductor QDs. 

The review is arranged as follows. Section \ref{problem} outlines the quantum control problem. Section \ref{method} provides a review of several popular techniques for coherent control, mainly based on a two-level quantum system. Section \ref{semi} discusses their applications in semiconductor QD spin systems. Finally in Section \ref{outlook}, we briefly summarize the review and outline some future directions for controlling semiconductors qubits.

\section{Quantum Control Problem}\label{problem}
\subsection{Basic control model}
We begin with an outline of the basic model in quantum control. The total Hamiltonian can be written in the form
 \begin{equation}\label{Hgeneral}
    H\left ( t \right ) =H_{s} +H_{c } \left ( t \right ), 
\end{equation}
where $H_{s}$ describes the static internal Hamiltonian of the qubit, $H_{c } \left ( t \right )$ is a time-dependent control Hamiltonian driving the logical gates to realize the desired dynamics. The control field may be an electric field \cite{Bayer2001}, a magnetic field \cite{Tiw2021}, an electromagnetic field \cite{Je2004}, or a combination of these \cite{Pf2017}.

For closed systems, the evolution of a quantum state $\left | \Psi(t) \right \rangle $ is described by the Schr\"{o}dinger equation
 \begin{equation}\label{}
 \frac{\partial }{\partial t} \left | \Psi(t)   \right \rangle =-\frac{i}{\hbar}{H} \left |\Psi(t)\right \rangle,
 \end{equation}
or equivalently, the density operator evolves according to the von Neumann equation
 \begin{equation}\label{}
  \frac{\partial }{\partial t}{\rho}(t)  =-\frac{i}{\hbar}[ {H}(t),  \rho(t)].
 \end{equation}
where $\hbar$ is the Plank constant. The solution is of the form $\rho (T) =U(T,0)\rho (0)U(T,0)^{\dagger } $ with the unitary operator
\begin{equation}\label{}
 U(T,0)=\mathcal{T}\exp\left [-\frac{i}{\hbar}\int_{0}^{T} H(t)\mathrm{d}t \right ],
\end{equation}
where $\mathcal{T}$ is the Dyson time-ordering operator.
The goal of quantum control is to find an appropriate $H_{c} \left ( t \right )$ to to produce either a state-state transition or enable an effective unitary operation, even in the presence of imperfections of experimental implementation and environmental noises.

\subsection{Fidelity measures for quantum state transfer and quantum gates}
To evaluate quantum control protocols, we introduce a few popular fidelity measures, and the goal of quantum control is to maximize such fidelity. 

The first class of control problems is the state-to-state transition. The fidelity measures the closeness between a target state $\rho_{\rm tar}$ and a final state $\rho(T)$ which evolves from a known initial state $\rho(0)$. More specifically, the state fidelity is defined as
\begin{equation}\label{}
  F_S=\operatorname{Tr}\left(\sqrt{\sqrt{\rho(T)}\rho_{\operatorname{tar}}\sqrt{\rho(T)}}\right).
\end{equation}
where $\sqrt{\rho}=\sum_m \sqrt{p_m}|m\rangle\langle m|$ denotes the square root of the positive operator $\rho=\sum_m p_m|m\rangle\langle m|$ with $p_m\geq0$ and $\sum_m p_m=1$, and Tr$[\cdot]$ denotes the trace of an operator.

The second class of control problems is to realize a target unitary operation corresponding to a quantum gate. Then the gate fidelity measures how close a quantum gate $U(T)$ is to the target gate $U_{\rm tar}$ \cite{Nielson2000}
\begin{equation}\label{}
F_{G}=\left|\frac{\mathrm{Tr}[U_{\rm tar}^{\dagger}U(T)]}{\mathrm{Tr}[U_{\rm tar}^{\dagger}U_{\rm tar}]}\right|^{2} .
\end{equation}
where $(\cdot)^{\dagger}$ denotes the Hermitian conjugate of an operator.
In multilevel systems, considering average fidelity of a unitary transformation of different initial states, another widely-used expression is \cite{JG2010}:
\begin{equation}\label{fide}
F=\frac{\operatorname{Tr}\left[U(T) U(T)^{\dagger}\right]+\left|\operatorname{Tr}\left[U_{\text {target }}^{\dagger} U(T)\right]\right|^{2}}{d(d+1)} .
\end{equation}
where $d$ is the dimension of the logical subspace and $U(T)$ here is the unitary operation obtained from control manipulation projected onto the logical subspace. Note that Eq. \eqref{fide} has taken it into account that the projection may not be a unitary operator and captures both computational and leakage errors.

 \section{Diverse methods of quantum control}\label{method}
 \subsection{Resonant excitation}
 Resonant excitation (RE) is the most commonly used control technique. For a simple two-state system, its Hamiltonian in the rotating-wave approximation reads
 \begin{equation}\label{Hre}
 {H}(t)=\frac{\hbar}{2}\left[\begin{array}{cc}
\Delta(t) & \Omega(t) \\
\Omega(t) & -\Delta(t)
\end{array}\right],
 \end{equation}
 where  the Rabi frequency $\Omega(t)$
describes the strength of the interaction between the quantum system and the external field, the detuning $\Delta(t)$ is the difference between the Bohr transition frequency $\omega _{12} $ and the carrier frequency $\omega$, each of which is generally time-dependent.

\begin{figure*}

  \includegraphics[width=6.0in]{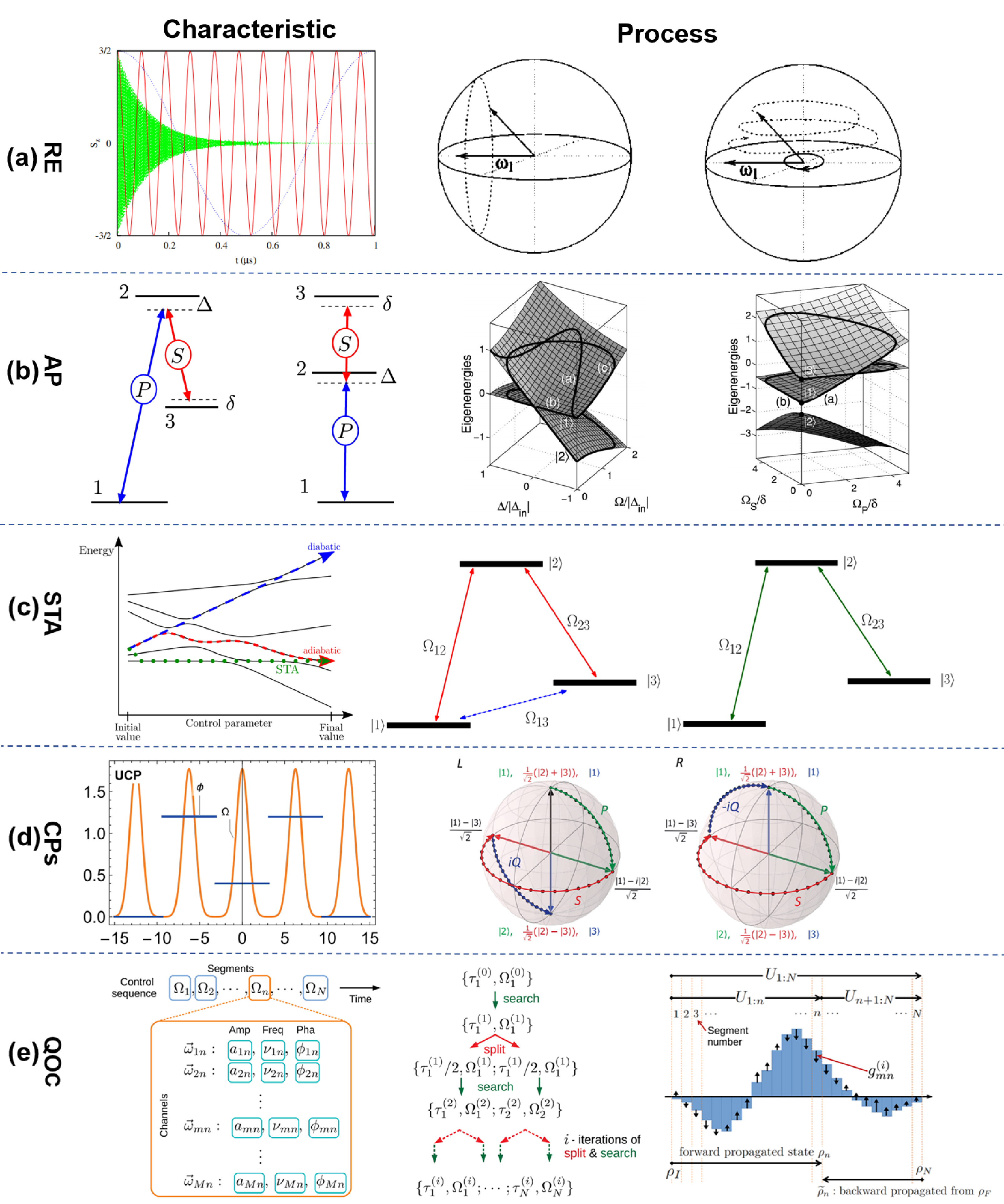}

   \caption{Diverse methods of quantum control with their characteristics and typical control processes. (a) Rabi oscillations at three different values
of the perpendicular magnetic field (left) (adapted from \cite{Bu2007}). A typical nutation of a qubit subject to a resonant transversal field illustrated on a Bloch sphere observed in the rotating frame (middle) and in the lab frame (right) (adapted from \cite{BT2020}). (b) The $\Lambda $-linkage pattern and the ladder pattern of energy levels in STIRAP (left) (Adapted from \cite{NV2017}). Typical surfaces of eigenenergies of a two-level system for RAP (middle) and STIRAP (right) (Adapted from \cite{LP2002}). (c) Schematic of adiabatic, diabatic, and STA processes (left), where along the STA path (green dots), the system evolves to the target state in a shorter time without necessarily traveling along the same level. Different strategies for STA in three-level systems, including the counterdiabatic STA (middle) with initial couplings (red, solid lines) and additional required STA coupling (blue, dotted line), and invariant-based inverse engineering STA (right) with modified STA couplings (green, solid lines) (adapted from \cite{DG2019}). (d) A diagram depicting the time dependence of the Rabi frequency and phase in CPs (left) (adapted from \cite{BT2021}). A typical illustration of the evolution for composite pulses sequence on a pseudo Bloch sphere with left (L) (middle) and right (R) (right) handedness (Adapted from \cite{To2020}). (e) A typical control structure for QOP (left), and the illustration of the gradients for amplitudes in the strongly modulating pulses direct search method (middle) and the GRAPE algorithm (right)  (Adapted from \cite{Mah2023}). For strongly modulating pulses direct search method, every time each segment is split into two equal parts, after searching the optimal control to minimize the cost functional.}
  
  \label{Methods}
  \end{figure*}

In RE, the carrier frequency of the radiation exactly matches the (constant) Bohr frequency of the transition, meaning $\Delta(t)= 0$. In this case, an observer in the rotating frame will see the quantum state simply precess about $\hat{x} $ on the Bloch sphere if the Rabi frequency is real [Fig. \ref{Methods}(a) middle]. The period of oscillation depends on the Rabi frequency, which is determined by the control filed coupling [Fig. \ref{Methods}(a) left]. And an observer in the lab frame will see the spin state spiral down over the surface of the Bloch sphere [Fig. \ref{Methods}(a) right]. The transition probability for resonant excitation is well known as
 \begin{equation}\label{}
 P=\sin ^{2}(\mathcal{A} / 2)=\frac{1}{2}[1-\cos (\mathcal{A})],
  \end{equation}
where the pulse area $\mathcal{A}=\int \mathrm{d} t \Omega(t)$ is often chosen to be an odd-integer multiple of
$\pi$ to produce a complete population inversion, or population swapping.

Resonant methods are constrained by the requirement that the detuning must vanish and the pulse area must be fixed exactly in practice. In experiment, the resonances may be broadened by inhomogeneities of the fxed field, and it can be suppressed by using molecular beam resonance method \cite{NF1950} with sharper formants.

\subsection{Adiabatic passage}
Adiabatic methods are based on the adiabatic theorem which guarantees that the system will follow the instantaneous eigenstate, provided the Hamiltonian varies sufficiently slowly \cite{TA2018}. These methods are more suitable for quantum state transfer than the gate operation. Without requiring precise control of pulse area and frequency detuning, the main advantage of adiabatic techniques is that they are more robust to pulse errors and system noise compared to conventional RE methods \cite{BD2009,GA2004}.

A well-known adiabatic control method is called adiabatic passage (AP), which mainly originated from two important techniques for transferring population between internal atomic/molecular levels, including the rapid adiabatic passage (RAP) \cite{NV2001,LA1987} and stimulated Raman adiabatic passage (STIRAP) \cite{KB1998,MP1997}.

\subsubsection{Rapid adiabatic passage}
The RAP technique is implemented in two-level atomic systems interacting with a chirped laser pulse. This technique is also called adiabatic following \cite{OJ1984}.

According to the adiabatic theorem, one can transfer the state vector in its Hilbert space to any desired superposition state by appropriately slowly (adiabatically) adjusting the Hamiltonian, when the initial state vector matches one of the instantaneous eigenvectors of the Hamiltonian.
Since the adiabatic dynamics of the process is determined by the topology of energy surfaces and can be completely predicted, the evolution of a state governed by time-dependent Schr\"{o}dinger equation is, thus, reduced to the topology of eigenenergies of multiple time-independent Schr\"{o}dinger equations \cite{LP2002,SG2001}.

To better understand the adiabatic process, we take a two-state system as an example and analyze its topology of eigenenergy surfaces. The Hamiltonian given in Eq. \eqref{Hre} is still valid \cite{LA1987}, while $\Omega(t)$ should be interpreted as a one-photon or multi-photon Rabi frequency with $\Delta_{0}(t)$ being the detuning from this resonance frequency. Additionally, another detuning $S(t)$ resulting from the difference of the dynamical Stark shifts associated with two energy levels should be taken into account. Then the detuning is replaced by
\begin{equation}\label{}
    \Delta(t)=\Delta_{0}(t)+S(t),
\end{equation}
where the second term refers to the effective dynamical Stark shift \cite{SC1997}. Note that the first term is dependent on the control field (time independent if the pulse produced by the control field remain unchanged) \cite{LP1999,TR2000}. Thus the variation of detuning (chirp) can be induced  either by direct sweeping of the frequency of one laser pulse, or the Stark shift due to additional fields.

The process can be completely described by the diagram of two surfaces:
\begin{equation}\label{}
   \epsilon_{ \pm}(t)= \pm\frac{1}{2} \hbar  \sqrt{\Omega(t)^{2}+\Delta(t)^{2}}
\end{equation}
which are also the the eigenvalues of
the Hamiltonian. The Hamiltonian in the adiabatic basis reads
\begin{equation}\label{}
H_{a}(t)=\hbar\left[\begin{array}{cc}
\epsilon_{+}(t) & -i \dot{\theta}(t) \\
i \dot{\theta}(t) & \epsilon_{-}(t),
\end{array}\right]
\end{equation}
where $\dot{\theta}(t)$ is the time derivation of the so-called mixing angle, which is defined as
\begin{equation}\label{}
\theta(t)=\frac{1}{2} \arctan \left[\frac{\Omega(t)}{\Delta(t)}\right],\quad-\frac{\pi }{2}\le  \theta\le 0.
\end{equation}
To satisfy the adiabatic condition and ensure the robustness, the process must be sufficiently far from energy crossings. More precisely, an eligible adiabatic evolution should guarantee that the nonadiabatic coupling (the off-diagonal elements) is much smaller than the difference between two eigenvalues, i.e.,
\begin{equation}\label{}
|\dot{\theta}(t)| \ll \epsilon_{+}(t)-\epsilon_{-}(t)=\sqrt{\Omega(t)^{2}+\Delta(t)^{2}}.
\end{equation}
In this case the state vector can remain aligned with an initial adiabatic state \cite{GS2005}. Fig. \ref{Methods}(b) (middle) depicts three different paths, denoted (a), (b) and (c), among which (a) correspond to direct chirping and (b) and (c) to Stark-chirped rapid adiabatic passage.

\subsubsection{Stimulated Raman adiabatic passage}
On the other hand, STIRAP is performed in a $\Lambda $-type three-level atomic system interacting with two radiation fields in a counterintuitive sequence [Fig. \ref{Methods}(b)]. It can efficiently transfer population between two discrete quantum states via an intermediate state, which is usually a radiatively decaying state \cite{HR1978}. Although STIRAP introduces an additional state, it is notably robust against the spontaneous emission from the intermediate state, along with other small variations of experimental conditions, such as laser intensity, pulse timing, and pulse shape, despite the fact that the duration of the radiative interaction may well exceed the radiative lifetime by several orders of magnitude \cite{Sh2013}.

Here we introduce the simplest version of STIRAP which allows a complete transfer of population along a three-state chain 1-2-3. The transfer starts from a populated quantum state 1 to a target quantum state 3, induced by two coherent radiation fields that couple the intermediate state 2 to states 1
and 3, labeled by the P and S respectively. Fig. \ref{Methods}(b) (left) shows the two common linkage patterns of radiative interactions for STIRAP. The Hamiltonian under RWA is typically written as \cite{Sh1990}
\begin{equation}\label{stirap}
 H(t)=\hbar\left[\begin{array}{ccc}
0 & \frac{1}{2} \Omega_{P}(t) & 0 \\
\frac{1}{2} \Omega_{P}(t) & \Delta & \frac{1}{2} \Omega_{S}(t) \\
0 & \frac{1}{2} \Omega_{S}(t) & \delta
\end{array}\right],
\end{equation}
where two detunings of adjacent states read
\begin{equation}\label{}
\begin{aligned}
&\hbar \Delta_{P}  = E_{2}-E_{1}-\hbar \omega_{P}=\Delta
\\&  \hbar \Delta_{S}= E_{2}-E_{3}-\hbar \omega_{S}=\Delta -\delta
\end{aligned}
\end{equation}
with $\omega_{P}$ and $\omega_{S}$ being the corresponding carrier frequency of the radiation field. STIRAP requires $\delta=0$, which means $\Delta=\Delta_{P}=\Delta_{S}$ for the $\Lambda $ linkage or $\Delta=\Delta_{P}=-\Delta_{S}$ for the ladder linkage.

One of the eigenvalues of
the Hamiltonian vanishes under the resonant condition ($\delta=0$). Diagonalizing Eq. \eqref{stirap} one can obtain three eigenstates
\begin{eqnarray}\label{}
\begin{aligned}
&|\Phi _{+} \rangle=\sin \Theta \cos \varphi\left|\phi_{1}\right\rangle-\sin \varphi\left|\phi_{2}\right\rangle+\cos \Theta \cos \varphi\left|\phi_{3}\right\rangle \\
&|\Phi _{-}\rangle=\sin \Theta \sin \varphi\left|\phi_{1}\right\rangle+\cos \varphi\left|\phi_{2}\right\rangle+\cos \Theta \sin \varphi\left|\phi_{3}\right\rangle \\
&|\Phi _{0}\rangle=\cos \Theta\left|\phi_{1}\right\rangle-\sin \Theta\left|\phi_{3}\right\rangle
\end{aligned}
\end{eqnarray}
with eigenvalues
\begin{eqnarray}\label{}
\begin{aligned}
&\lambda_{ \pm}=\frac{\hbar}{2}\left(\Delta  \pm \sqrt{\Delta ^{2}+\Omega _{P}^{2}+\Omega _{S}^{2}}\right) \\
&\lambda_{0}=0,
\end{aligned}
\end{eqnarray}
where $\phi _{k}\ (k=1,2,3)$ are the wave functions of the
unperturbed states of the $\Lambda $-system and the mixing angles $ \Theta$ and $\varphi$ are given by
\begin{eqnarray}\label{}
\begin{aligned}
&\tan \Theta  =\frac{\Omega_{P}(t)}{\Omega_{S}(t)} \\
&\tan 2 \varphi(t)=\frac{\sqrt{\Omega_{P}(t)^{2}+\Omega_{S}(t)^{2}}}{\Delta}.
\end{aligned}
\end{eqnarray}
By smoothly varying the tunneling rates, corresponding to the mixing angle $\Theta$ evolving from $0$ to $\pi/2$, one can adiabatically transfer the state $\phi _{1}$ to $\phi _{3}$, i.e., initially $\left|\Omega_{S}(t)\right|>0 \text { while } \Omega_{P}(t)=0 \text { or }\left|\Omega_{S}(t)\right| \gg\left|\Omega_{P}(t)\right|$, and at the end $\left|\Omega_{P}(t)\right|>0 \text { while } \Omega_{S}(t)=0 \text { or }\left|\Omega_{P}(t)\right| \gg\left|\Omega_{S}(t)\right|$. Fig. \ref{Methods}(b) (right) depict two possible paths, which correspond to the intuitive ($\left | 1  \right \rangle $ to $\left | 2  \right \rangle $) and counterintuitive ($\left | 1  \right \rangle $ to $\left | 3 \right \rangle $) pulse sequences respectively.

\subsection{Shortcuts to adiabaticity}
As discussed in the last subsection, adiabatic passage techniques require sufficiently slow manipulation of quantum system, which indicates the timescale of evolution is relatively long. However, undesirable environmental interactions may cause dissipation and dephasing in quantum systems and lead to a degraded performance of quantum control. In order to accelerate the conntrol process, especially within the  timescale of decoherence time, a type of method called shortcuts to adiabaticity (STA) has been proposed, which can provide a faster route to reach the final states compared to the adiabatic methods [Fig. \ref{Methods}(c) left]. The concept of STA was first explicitly coined in 2010 \cite{ch2010}. Since then, STA methods and their applications have been experiencing rapid development \cite{DG2019}. Now it has become a field of its own, but some STA methods are related to or partially overlaping with other control methods \cite{Wu2017,Ab2015}.

Various STA methods have been well developed, such as inverse engineering approaches, invariant and scaling laws based engineering, variational methods, fast forward approaches, fast quasiadiabatic (FAQUAD), STA methods via optimal control theory \cite{Be2009,Ka2017,Le1982,Ma2018,Ka1995,Ko2017}. The most widely used one is the counterdiabatic driving, where an extra counterdiabatic Hamiltonian is added to the original one. By doing so, the dynamics follows exactly the approximate adiabatic evolution driven by the original Hamiltonian. Hence, the system can evolve like the eigenstates of the original (reference) Hamiltonian, even for very fast evolution under the influence of the total Hamiltonian. Here we show a simple example of counterdiabatic driving in a two-level system.

We still start from Eq. \eqref{Hre}. To achieve the shortcut, an additional counterdiabatic Hamiltonian can be introduced with the form
\begin{equation}\label{}
H_\mathrm{C D}(t)=\frac{\hbar}{2}\left(\begin{array}{cc}
0 & -i \Omega_{a}(t) \\
i \Omega_{a}(t) & 0
\end{array}\right),
\end{equation}
with $\Omega_{a}(t)=\left[\Omega(t) \dot{\Delta}(t)-\dot{\Omega}(t) \Delta(t)\right] / [\Delta^{2}(t)+\Omega^{2}(t)]$. Then, in the adiabatic basis of the original Hamiltonian, the transformed shortcut Hamiltonian $H'(t)=H_{0}(t)+H_\mathrm{C D}(t) $ can be diagonalized as
\begin{equation}\label{}
H_{a}^{\prime}(t)=\hbar\left[\begin{array}{cc}
\epsilon_{-}(t) & 0 \\
0 & \epsilon_{+}(t)
\end{array}\right] .
\end{equation}
Thus, adiabatic passage can be more effective in population transfer, because any transition in the adiabatic basis of the original Hamiltonian is suppressed. However, this evolution is no longer adiabatic, because it does not coincide with the eigenstate of the total Hamiltonian. Hence, the term ``shortcuts to adiabaticity'' is not always appropriate.

Dynamical invariants and invariant-based engineering constitute another major route to design STA protocols. For a given Hamiltonian there are many possible invariants. These invariants satisfy:
\begin{equation}\label{}
\frac{d I}{d t}=\frac{\partial I}{\partial t}+\frac{i}{\hbar}[H, I]=0.
\end{equation}
Under the adiabatic condition, which means long operation times, Eq. (22) becomes $[H, I] \approx 0.$ Sometimes we only demand that the invariant and the Hamiltonian commute at the start and end of the process, i.e., $[I(0), H(0)]=\left[I\left(t_{f}\right), H\left(t_{f}\right)\right]=0$. Thus the eigenstates of the invariant and the Hamiltonian coincide at initial and final times. If no level crossings take place, the final state will keep the initial populations. One could design the invariant according to the ideal state transfer. For example, consider a two-level system 
\begin{equation}\label{}
H(t)=\frac{\hbar}{2}\left(\begin{array}{cc}
\Delta(t) & \Omega_{R}(t)-i \Omega_{I}(t) \\
\Omega_{R}(t)+i \Omega_{I}(t) & -\Delta(t)
\end{array}\right),
\end{equation}
whose two eigenvectors are given by
\begin{equation}\label{}
\begin{array}{l}
\left|\phi_{+}(t)\right\rangle=\binom{\cos (\theta / 2) e^{-i \varphi  / 2}}{\sin (\theta / 2) e^{i \varphi / 2}}, \\
\left|\phi_{-}(t)\right\rangle=\binom{\sin (\theta / 2) e^{-i \varphi  2}}{-\cos (\theta / 2) e^{i \varphi / 2}},
\end{array}
\end{equation}
we can set the invariant in the most general form:
\begin{equation}\label{}
I(t)=\frac{\hbar}{2}\left(\begin{array}{cc}
\cos [\theta(t)] & \sin [\theta(t)] e^{-i \varphi (t)} \\
\sin [\theta(t)] e^{i \varphi (t)} & -\cos [\theta(t)]
\end{array}\right).
\end{equation}
On the basis of Eq. (22), $\theta (t)$ and $\varphi (t)$ must satisfy:
\begin{equation}\label{}
\begin{array}{c}
\dot{\theta}=\Omega_{I} \cos \varphi -\Omega_{R} \sin \varphi, \\
\dot{\varphi}=\Delta-\cot \theta\left(\Omega_{R} \cos \varphi+\Omega_{I} \sin \varphi\right).
\end{array}
\end{equation}
Suppose a particular solution to the Schrödinger equation is
\begin{equation}\label{}
|\psi(t)\rangle=\left|\phi_{+}(t)\right\rangle e^{-i \gamma(t) / 2},
\end{equation}
we can retrieve the control parameters in terms of the auxiliary functions,
\begin{equation}\label{}
\begin{array}{c}
\Omega_{R}=\cos \varphi  \sin \theta \dot{\gamma}-\sin \varphi \dot{\theta}, \\
\Omega_{I}=\sin \varphi \sin \theta \dot{\gamma}+\cos \varphi \dot{\theta}, \\
\Delta=\cos \theta \dot{\gamma}+\dot{\alpha} .
\end{array}
\end{equation}
Then different state manipulations can be realized with appropriate boundary conditions, for example, $\theta(0)=0 $ and $\theta\left(t_{f}\right)=\pi$ imply perfect population inversion at a time $t_{f}$.

Note that the counterdiabatic driving and invariant-based approach generally lead to different control protocols. Considering a three-level $\Lambda $ system consisting of two “ground levels” $\left | 1 \right \rangle $ and $\left | 3 \right \rangle $ and a central excited level $\left | 2 \right \rangle $ coupled with time-dependent terms $\Omega _{12} $ and $\Omega _{23} $. If we apply the counterdiabatic STA technique directly to this $\Lambda $ system, we need an additional coupling between the two levels $\left | 1 \right \rangle $ and $\left | 3 \right \rangle $ [Fig. \ref{Methods}(c) middle]. On the other hand, the invariant-based engineering that may or may not populate level $\left | 2 \right \rangle $ have no need for an additional coupling between  $\left | 1 \right \rangle $ and $\left | 3 \right \rangle $ [Fig. \ref{Methods}(c) right].

\subsection{Composite pulses}
Composite pulses (CPs) is a powerful method that combines robustness to experimental fluctuations and high control accuracy without a complicated process. Using a sequence of pulses with specific computed phases, one can compensate the errors induced by control process. For example, off-resonant detunings can also be exploited as control parameters \cite{EK2019}. Fig. \ref{Methods}(d) (middle and right) illustrate typical evolution on a Bloch sphere by using a sequence of resonant pulses with total area of just $2\pi$ for a left-handed and right-handed system respectively.

Previous studies have demonstrated the generality and the universality of CPs approach to realize robust population inversion or implement robust quantum gate \cite{TI2011,XW2014,GT2014}. Usually these various CPs scheme consider only one source or several specific kinds of noises, mainly imperfections arising from the quantum system or the external field \cite{GH2016,XW2014}. Here we introduce a relatively general type of CPs named universal composite pulses \cite{Ds2013} in a two-level system as an example  [Fig. \ref{Methods}(d) left].

Under the premise that the CPs duration is shorter than the decoherence times in the system, the evolution of the system can be
described by a propagator $U$, which is conveniently parameterized with the three
real $\mathrm{St\ddot{u} ckelberg} $ variables $q \ (0 \le  q \le  1)$, $\alpha$ and $\beta$ as
\begin{equation}\label{}
    U=\left[\begin{array}{cc}
q e^{i\alpha} & p e^{i \beta} \\
-p e^{-i \beta} & q e^{-i \alpha}
\end{array}\right],
\end{equation}
where $p=\sqrt{1-q^{2}}$. Then $p^{2} $
represents the transition probability of states. An additional phase to $\beta$ will arise if a constant phase shift is introduced to the Rabi frequency, i.e., $\Omega(t) \rightarrow \Omega(t) e^{i \phi}$, and the propagator becomes
\begin{equation}\label{}
U(\phi)=\left[\begin{array}{cc}
q e^{i\alpha} & p e^{i(\beta+\phi)} \\
-p e^{-i(\beta+\phi)} & q e^{-i \alpha}
\end{array}\right].
\end{equation}
We keep other parameters unchanged in a composite sequence of $n$ pulses each with a phase $\phi_{k} \ (k=1,2,\cdots,n)$, and the whole evolution propagator reads
\begin{equation}\label{}
U^{(\mathrm{total})}=U\left(\phi_{n}\right) \cdots U\left(\phi_{2}\right) U\left(\phi_{1}\right).
\end{equation}
If our objective is to implement a NOT gate with robustness to deviations in all pulse parameters, we should minimize the
probability for no transition $\left|U_{11}^{(\mathrm{total})}\right|^{2}$. Making an assumption that the constituent pulses are identical apart from their phases, we can make the error term arbitrarily small in a neighborhood of $q=0$, such that these segments are approximately $\pi$-pulses. For instance, for a five-pulse sequence,
\begin{equation}\label{}
\begin{aligned}
U_{11}^{(5)}= & \left\{\left[1+2 \cos \left(2 \phi_{2}-\phi_{3}\right)\right] e^{i\alpha}+2 \cos \left(\phi_{2}-\phi_{3}\right) e^{-i \alpha}\right\} q \\
& +O\left(q^{3}\right)
\end{aligned}.
\end{equation}
The error term vanishes if we choose appropriate phases: $\left\{\phi_{2}=5 \pi / 6\right. ,\  \left.\phi_{3}=\pi / 3\right\} $ and $\left\{\phi_{2}=11 \pi / 6, \ \phi_{3}=\pi / 3\right\} $.

\subsection{Quantum optimal control}
 Classical optimal control theory aims at designing the best control for a given target while meeting certain conditions and obeying Pontryagin's maximum principle. The optimal control theory is a long-standing technique with a broad range of applications in a variety of systems, initially developed to solve problems related to space rage \cite{Po1962}. Then quantum optimal control (QOC) theory emerged to extend the control problem into the realm of quantum theory \cite{BL1973}. The QOC was widely used in in the nuclear magnetic resonance community for designing shaped radio-frequency pulses with certain spectral properties during the 1970s. Recently there is emerging interest in using the QOC techniques to realize high-fidelity quantum control for quantum technologies. 

 In the quantum context, the task is usually to find a ``control field'' that steers the time evolution of a quantum system in some desired way, including realizing ideal state transfer and quantum gate operation in the minimum possible time or with maximum fidelity. QOC formalizes this problem and give numerical solutions to it. To do so, the QOC algorithm either makes use of a simulated model or adaptive learning from the experiment, and we refer them to open-loop and close-loop quantum optimal control respectively.

 Ignoring the unknown part of Hamiltonian related to environment in Eq. \eqref{Hgeneral} (unless we want to see its effects or try to minimize them), the evolution of the system is determined by the control field $ \left \{ \omega _{m}(t) \right \} $. We often assume the Hamiltonian to be linear under controls,
 \begin{equation}\label{CP1p}
 H(t)=H_{s}+\omega _{1}(t) H_{1}+\omega_{2}(t) H_{2}+\cdots.
 \end{equation}
To facilitate the numerical simulation, we shall discretize the entire control field of duration $T$ into $N$ piecewise constant segments $\left \{ \left \{ \omega _{m1}  \right \} , \left \{ \omega _{m2}  \right \},\left \{ \omega _{m3}  \right \},\dots \left \{ \omega _{mN}  \right \} \right \}$ with distinct controls keeping stable in each fragment. Note that non-linear couplings may also occur, and parameters in addition to amplitude, e.g., frequency and phase, can be likewise set as variables. Collectively, we could represent the control as a vector $ \vec{\omega} _{mn} $ in the multi-parameter case with Hamiltonian $H_{m}^n(\vec{\omega}_{m n})$ [Fig. \ref{Methods}(e) left]. Thus the control Hamiltonian for the $n$th segment is of the form
\begin{equation}\label{CPn}
 H_{c}^{n} =\sum_{m=1} {H}_{m}^{n}\left(\vec{\omega}_{m n}\right).
\end{equation}
The corresponding Hamiltonian and propagator for the $n$th segment of duration $\tau_{n}$ are
\begin{equation}\label{}
    H^{n}=H_{s}+H_{c}^{n} , \quad U_{n}=e^{-i H_{n} \tau_{n}}.
\end{equation}
And the overall propagator is simply given by
\begin{equation}\label{}
 U(0, T)=U_{1: N}=U_{N} U_{N-1} \cdots U_{n} \cdots  U_{2} U_{1}.
\end{equation}

To mathematically apply QOC methods, the goal here is to minimize a cost functional
 \begin{equation}\label{}
 J=\Phi[\left \{ \mathbf{\phi }_{k} (T) \right \} ]+\int_{0}^{T} L\left[ \left\{ \mathbf{\phi }_{k} (t) \right \}, \left\{ {\vec{{\omega}}_{m} }  (t) \right \}\right] \mathrm{d} t,
 \end{equation}
where $\Phi$ and $L$ are the terminal cost and running cost respectively, and $\left \{ \phi _{k} (t) \right \} $ are the time-evolved initial states $\left \{ \phi _{k}  \right \} $ under controls. $\Phi$ is the main part of the functional, and for fixed duration $T$ and target quantum state, we can usually adopt a simple expression like $\Phi =1-F$ with $F$ being the target fidelity we want to maximize, while assuming $L=0$. If the time and energy cost are taken into account, we need to build an appropriate model for $L$. For example, the choice $\Phi=0, \ L=1$ minimizes the transfer time $T$.

Before getting into details of specific QOC methods and algorithms, we would like to add some more general criteria. First, we need to set a stopping condition, which often requires the optimization functional to be restricted to a certain range. Then, an initial guess of control fields is necessary. A good guess will significantly speed up calculations. Besides, for practical considerations, we should sometimes specify the amplitude and the limits of amplitude variations (upper limit and lower limit) based on parameters of devices used in experiments, e.g., the capacity, sampling rate and model resolution.

Over the last two decades, numerous QOC methods
have been proposed, including direct search, gradient method, variational method, asymptotic method and machine learning method, etc. The first three methods are rather popular and have been successfully applied to many quantum information processing tasks.
Direct search method relies on cleverly sampling
the search space starting from a random initial
guess [Fig. \ref{Methods}(e) middle]. It is not as accurate as the gradient method, but it has a larger radius of convergence. On the other hand, the gradient method concentrates on local optimum by evaluating local gradients. It can provide a result with high accuracy and assure that the obtained solution satisfies the necessary conditions, but the radius of convergence is usually small.

Here we take the celebrated Gradient Ascent Pulse Engineering (GRAPE) algorithm as an example [Fig. \ref{Methods}(e) right]. It starts from a random initial sequence of $N$ segments, each having the same duration $\tau $. We assume that only amplitude is accepted as the control parameter, so for the $n$th segment, ${H}_{m}^n={H}_{m}$. This is a valid assumption and the most common condition, since it is easier to execute in both theory and experiment. Accordingly, similar to Eq. \eqref{CPn}, the control Hamiltonian reads ${H}_{c}^{n} =\sum_{m=1}^{} \omega_{m n} {H}_{m}$. In the $i$th iteration, each segment is corrected according to
  \begin{equation}\label{}
\omega_{m n}^{(i)}=\omega_{m n}^{(i-1)}+\epsilon \tau g_{m n}^{(i)},
\end{equation}
where $\epsilon $ is the step size and $g_{m n}^{(i)}$ is the first-order gradient with simple analytical expression.

Apart from these common methods, machine learning method is finding applications everywhere thanks to the ever-increasing computational power. Meanwhile, hybrid algorithms which combine the strengths of the original methods have arisen. These methods may play an important role in designing complex and robust control sequences for larger systems with severe parameter constraints.

\begin{figure*}

  \includegraphics[width=6.0in]{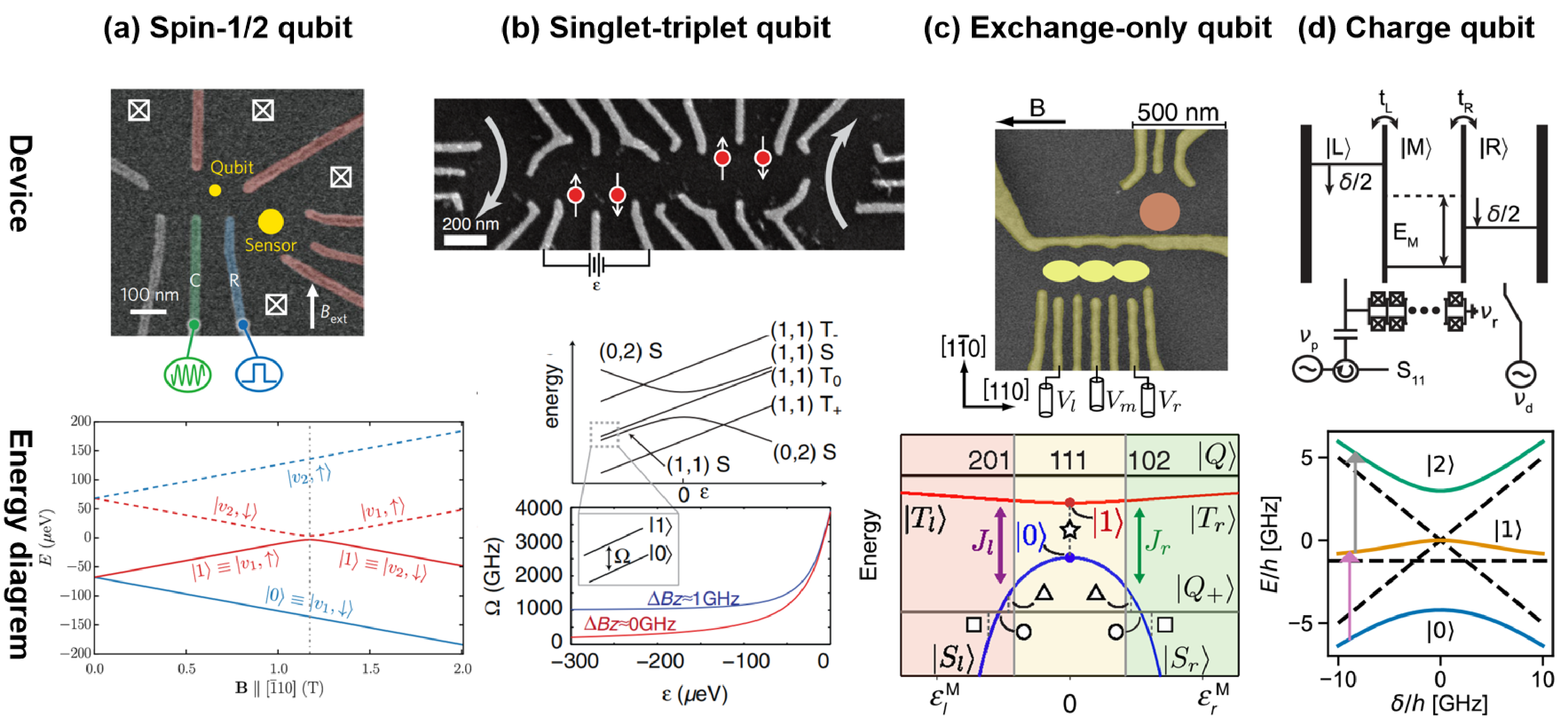}

  \caption{The four major qubit types covered in this review, with typical device images and energy-level diagrams. (a) The device layout of a quantum-dot electron spin-1/2 qubit (adapted from \cite{YJ2018}), and a typical image about energy levels of a silicon quantum dot in a magnetic field B (adapted from \cite{Bo2018}). (b) Scanning electron micrograph of a two-qubit device in GaAs quantum dots, and a energy level diagram showing the two-electron spin states of the double quantum dot and energy splitting(adapted from \cite{Ni2017}). (c) False color micrograph of a quantum-dot-based resonant exchange qubit and the corresponding energy level diagram (adapted from \cite{MJ2013}). (d) Schematic diagram of a triple-dot device for a charge quadrupole qubit and spectrum of the Hamiltonian for equal tunnel couplings as a function of detuning $\delta $(adapted from \cite{Kr2021}).}
  \label{qubits}
  \end{figure*}
\section{Quantum Control In semiconductor Qubits}\label{semi}
\subsection{Overview of semiconductor quantum dot qubits}

There are multiple ways to encode quantum information in semiconductor QDs. For electrons and holes, both spin and charge degrees of freedom can be employed as qubits \cite{ZX2017}. We begin this section with a brief introduction to these qubits.

With respect to the spin degree of freedom, Loss-Divincenzo spin-1/2 qubits, singlet-triplet qubits and exchange-only qubits have been successively proposed and realized in experiments. First, it is natural to consider a single localized electron or nucleus spin as a qubit. Once an electron or nucleus is put into a magnetic field, the energy levels of spin-up and spin-down are split by the so-called Zeeman energy. This two-level system can be used as a spin-$1/2$ qubit, distinguished from other types of spin qubits [Fig. \ref{qubits}(a)]. To manipulate this type of qubit, microwave (MW) bursts via an antenna were used to generate an oscillating magnetic field \cite{HS2018,MJ2014}. This approach is called electron spin resonance (ESR) for controlling electron spins, or nuclear magnetic resonance (NMR) for controlling nuclear spins.

Another type of spin qubit, the singlet-triplet qubit, is encoded in the singlet-triplet spin subspace of two electrons trapped in a double QD  \cite{JP2005} [Fig. \ref{qubits}(b)]. Generally, the encoded states are $S$ and $T_{0}$, and the effective control Hamiltonian can be written as
\begin{equation}\label{}
    H_{S T}=J\left(\varepsilon_{s}\right) \frac{\sigma_{z}}{2}+\Delta E_{z} \frac{\sigma_{x}}{2},
\end{equation}
where the energy of exchange splitting of encoded states $J\left(\varepsilon_{s}\right)$ depends on the detuning $\varepsilon_{s}$ representing the electrochemical potential difference, and $\Delta{E}_{z}$ is the Zeeman energy difference of two spins. The evolution of the system depends on the size relationship of these two parameters, and qubit operations can be implemented by tuning the Heisenberg exchange interaction between the two spins. This manipulation process can be entirely electrically driven, making it both operationally convenient and compatible with existing semiconductor electronics.

A quite distinctive scheme different from above qubits is to implement a qubit solely by exchange interaction without a Zeeman energy difference. This idea leads to the exchange-only qubit, which is encoded by  the spins of three electrons in a triple QD [Fig. \ref{qubits}(c)]. By appropriate control of the exchange couplings between dots, one can perform any qubit rotation in theory. A great advantage of the exchange-only qubit is that it enables fast all-electrostatic operations while avoiding  the complexities of magnetic field control.

In experiments, spin qubits were first demonstrated in in GaAs/AlGaAs heterostructures for their relatively simple fabrication \cite{JP2005} . Besides, the GaAs platform has a single conduction band valley with a small effective mass, which results in favourable electrical conductivity. For such III–V heterostructures, hyperfine interaction with the lattice nuclear spins is the dominant source of qubit decoherence. Thus dynamic nuclear polarization and dynamical decoupling sequences have been developed to suppress the noise from nuclear spin baths \cite{Fo2009,Ma2017}. Then since the first demonstration of silicon spin qubits in 2012 \cite{Pl2012}, interests have been refocused on the group IV material systems with low concentrations of nuclear spins. Moreover, nuclear spins can be further reduced through isotopic purification, further increasing the qubit coherence time. There are also emerging interests in  SiGe/Ge/SiGe quantum-well heterostructures confining high-mobility hole gases \cite{Scappucci2021} , where quantum dots can be formed to encode hole spin qubits.

Besides the spin degree of freedom, charge qubits are also of interest. Their states can be easily defined by the electron occupation in QDs and the readout and manipulation could be rather straightforward using electron devices [Fig. \ref{qubits}(d)]. This leads to a hybrid qubit that combines the merits of both: the long coherence time of spin qubits and the direct manipulation with short operation time of charge qubits.

\subsection{Practical aspects in controlling semiconductor qubits}
\begin{figure*}\centering
	\includegraphics[width=15cm]{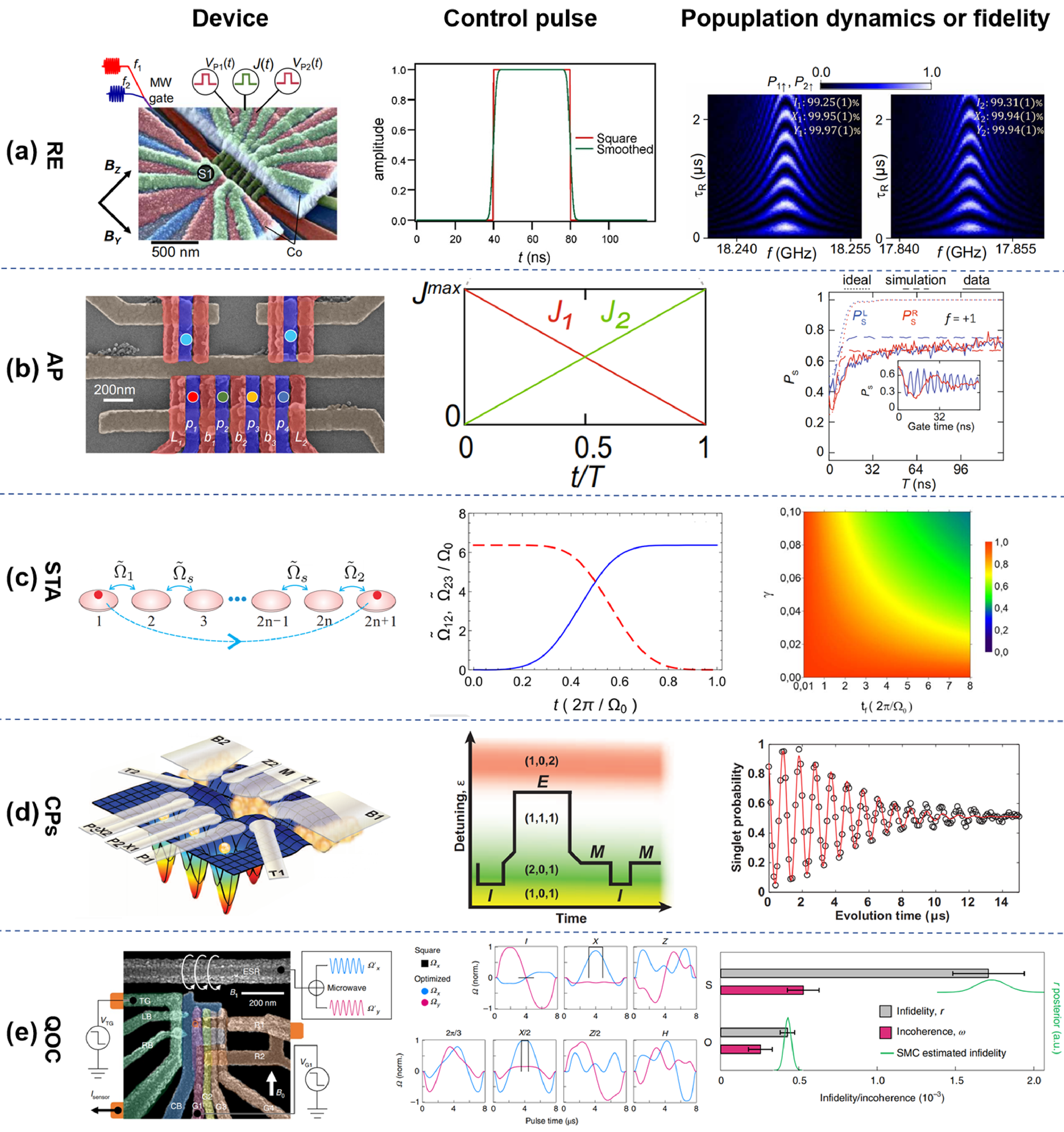}

	\caption{Applications of diverse control methods to semiconductor qubits: device diagrams, control pulses and gates fidelity (or population dynamics). (a) False-color scanning electron microscope image of a spin-based two-qubit silicon quantum processor (left); Comparison of a common control square pulse and a smoothed pulse to simulate the actual environmental conditions; Spin-up probabilities as a function of drive frequency $f$ and microwave burst length $\tau_{\mathrm{R}}$ when driven on resonance for each qubit (right). (adapted from \cite{Mi2022}) (b) False-color scanning electron micrograph of a quadruple quantum-dot device for adiabatic quantum-state transfer of both single-spin eigenstates and two-spin singlet states (left); Change in exchange-coupling strengths between qubits for the adiabatic quantum-state transfer step (middle); Singlet return probabilities of the left and right pairs as a function of interpolation time $T$ (right). (adapted from \cite{Ka2021}) (c) Direct transfer of one electron in an array of 2n + 1 QDs (left); Typical STA pulse shape (middle) with $\Omega _{12} $ (solid, blue), $\Omega _{23} $ (dashed, red); Fidelity $F$ versus dephasing rate $\gamma $ (in units of $5 \times 10^{7} s^{-1}$) and the operation time $t_{f}$ for a triple QD by inverse engineering STA (right). (adapted from \cite{Yb2018}) (d) Schematic diagram of a triple-dot device depicting the gate layout and the resulting electrostatic control of the potential landscape (left); The control composite pulses within all-electrical operation to measure charge noise dynamics in the system (middle); Sveraged double-dot singlet-triplet oscillations at a magnetic field of 0.1 $T$ (right). (adapted from \cite{En2015}) (e) Scanning electron micrograph image of a SiMOS qubit device (left); Microwave modulation $\Omega _{x} $ (blue), $\Omega _{y} $ (red) for the basic 7 types of Clifford gate through GRAPE iteration (middle); The infidelity ($r=1-F$) and incoherence ($\omega$) for sequences using square pulses (scheme S) and GRAPE optimised pulses (scheme O) in the randomised benchmarking result (right). (adapted from \cite{YC2019})}
  \label{aplfig}
\end{figure*}

$\mathit{Resonant \; excitation} $\quad The principle of resonant excitation has been applied in almost all cases except for adiabatic methods. Apart from the common ways to manipulate spin-$1/2$ qubits based on RE, such as the ESR and NMR type, another approach, known as electric-dipole spin resonance (EDSR) \cite{ZD2018}, apply a magnetic field gradient to modify the effective magnetic field experienced by the electron under electric driving, by means of spin-orbital coupling (SOC) of the semiconductor or an integrated
micromagnet [Fig. \ref{aplfig}(a)]. Thus, electron in this environment can feel an effective oscillating magnetic field if it is driven by an oscillating electric field. Therefore, MW bursts can be applied directly on a single electrode and the effective oscillating magnetic field is proportional to its voltage amplitude \cite{ZX2018}.

~\\

$\mathit{Adiabatic  \; passage} $\quad Gate operations in semiconductor qubits are mainly realized through coherent spin manipulation in QDs via electric, magnetic, and spin-orbit coupling control. Since adiabatic passage techniques have demonstrated its potential to achieve high-fidelity operations in the presence of small variations of environmental parameters, adiabatic quantum information processing in arrays of spin qubits has become the focus of theoretical research \cite{BD2009,SV2007,OS2013,BY2019}.

Some works have already realized the transfer of spin states in semiconductor qubits via electron spin shuttling \cite{Fu2017,Fl2017}. Furthermore, they successfully extended it to multi-level systems by shuttling an entangled state to distant sites through an artificial spin chain. Using a simple, linear ramp of the detuning energy in an array of QDs with a magnetic field gradient, the system undergoes an adiabatic transition at each energy crossing, with one party of an entangled spin pair being swapped with the next spin \cite{Na2018}. Moreover, recent studies proposed a new scheme to transfer both single-spin eigenstates and two-spin singlet states in a GaAs quadruple quantum-dot device without involving the physical motion of electrons [Fig. \ref{aplfig}(b)] \cite{Ka2021}. Fig. \ref{aplfig}(b) (middle) illustrate the change in exchange-coupling strengths between qubits during the adiabatic transfer process from spin 1-3. And Fig. \ref{aplfig}(b) (right) shows the results of the gradually rising return probabilities for the left and right pair of which we want to swap spin states. The experimental data suggests successful adiabatic transfer.   

As for STIRAP, since it was originally designed for atomic systems \cite{KB1998}, it is an ideal scheme at least in theory for quantum control in the semiconductor QDs because of their atomic-like density of states \cite{UH2006}. Although STIRAP has yet to be demonstrated experimentally in QDs, several important steps have been made \cite{Xu2008}. On the theoretical side, it has attracted significant interest, with numerous schemes proposed as follows.

One of the inchoate proposals is to use  optical excitations (excitons) in two coupled QDs as qubits. One can construct unconditional and conditional quantum gates by means of Coulomb interactions between the optically excited electrons and holes \cite{Ho2000}. Then they found that voltage control may be a better choice compared to optical control. With electrons in a narrow quantum well confined using external gate voltages, this qubit is represented by the spin of two electrons in a vertically  tunnel-coupled double QD structure. Thanks to strong Coulomb correlation effects, the carrier states are substantially renormalized, which can be used for entanglement detection, transport and disentanglement measurement \cite{Tr2003,Ho2006}.

Another scheme based on the fact that a single semiconductor QD inside a microcavity can create single and entangled photons in the presence of a lateral electric field. By utilizing this cavity-assisted STIRAP, one can promote the surplus electron from the ground state to the excited state, via excitation of a pump pulse and optical coupling to the charged exciton \cite{Ja2011}. Additionally, coherent manipulation of an asymmetric double-QD structure \cite{Vt2007} or electron transfer between the
ground states of two coupled QDs is an alternative method \cite{Fo2013}.
There are also quite a few studies examining the measurement, spin transport, states preparation and robustness in STIRAP processes \cite{Pa2001,Fa2005,Br2001,Ko2013}.
~\\

$\mathit{Shortcuts \; to \; adiabaticity} $\quad The working principle of STA has been
demonstrated experimentally in many other platforms, such as cold
atoms, Bose-Einstein condensate in
an optical lattice, nitrogen-vacancy centers, and trapped ions \cite{Yx2016,Mg2012,Jz2013}. Although it is technologically challenging for experimental realizations in semiconductor qubits, quite a few STA protocols
have been proposed to overcome the long run-time while providing robustness associated with adiabatic evolutions. Ban Y. et al. put forward schemes applied for fast and robust control of electron spin qubits in a single \cite{Yb2012} and a double QD \cite{Yb2014}, and for electron transfer and spin entangled
state in a long QD array \cite{Yb2018,BY2019} [Fig. \ref{aplfig}(c)].
A STA protocol based on invariant-based engineering to realize a nonadiabatic electron transfer was used in the work of Masuda S. et al. They showed its robustness in contrary to the normal $\pi$ pulse protocol and possibility to be extended to multi-electron systems to implement two-qubit gates \cite{Sh2018}.

As for quantum gate operation, Li Y.-C. et al. designed a qubit encoded in the electron spin among two QDs that can simultaneously perform an arbitrary four-dimensional qubit rotation \cite{Li2018}. And Yan T. et al. pointed out the possible experimental realization of nonadiabatic shortcut to non-Abelian geometric gates in QDs \cite{To2019}. Besides, a recent research demonstrated that a FAQUAD approach achieved higher gate fidelity compared to other protocols \cite{Df2022}. Moreover, Xu L. et al. presented a real-time simulation for the dynamics of braiding a pair of Majorana zero modes via a QD that can be performed with a STA approach \cite{Lu2023}.

~\\

\begin{table*}[t!]
\centering
\caption{An overview of the five quantum control methods discussed in the review.}
\label{comparison}
\begin{tabular}{lll}
\hline
Method              & Pros and cons                                                                                                                                                                  & Practical aspects in semiconductor qubits                                                                                                                                                                                   \\ \hline
\\ Resonant excitation   & \begin{tabular}[c]{@{}l@{}}Pros: simple, widely used\\ Cons: strict resonant condition
\end{tabular}                                                                                                                                                  & \begin{tabular}[c]{@{}l@{}}Manipulation:\\ ESR by engineered magnetic fields \cite{Ko2006,MJ2014,Ve2014}\\ EDSR via natural spin–orbit fields \cite{Nad2010,Cor2018,Cri2018}  \\ \ \ \ \ \ \ \ \ \ \ \ \ \ \ synthetic spin–orbit fields \cite{Ka2014,Za2018} \end{tabular}                                                                          \\ \\
\\
Adiabatic passage   & \begin{tabular}[c]{@{}l@{}}Pros: insensitive to variations of\\ \ \ \ \ \ \ \ \ control parameters \\
Cons: slow \end{tabular}                                                                             & \begin{tabular}[c]{@{}l@{}}Experimental advances:\\ RAP: realization of spin states transfer \\ \ \ \ \ \ \ \ \ via electron spin shuttling \cite{Fu2017,Fl2017}\\ \ \ \ \ \ \ \ \ spin states transfer in multi-level systems \\ \ \ \ \ \ \ \ \ through spin chain \cite{Na2018} \\ \ \ \ \ \ \ \ \ two-spin singlet states transferin a GaAs \\ \ \ \ \ \ \ \ \ quantum-dot device \cite{Ka2021}\\ STIRAP: demonstration of certain steps  \cite{Xu2008} \\ \ \ \ \ \ \ \ \ \ \ \ \ \ \ excitons in two coupled QDs \cite{Ho2000} \\ \ \ \ \ \ \ \ \ \ \ \ \ \ \  cavity-assisted STIRAP\cite{Ja2011}\end{tabular}\\ \\ \\
Shortcuts to adiabaticity                 & \begin{tabular}[c]{@{}l@{}}Pros: flexible;\\ \ \ \ \ \ \ \ \ faster than adiabatic passage\\
Cons: relatively high control complexity \ \ \  \end{tabular}           & \begin{tabular}[c]{@{}l@{}}Several protocols:\\ Fast and robust control of electron spins \cite{Yb2012,Yb2014,Yb2018,Yb2019}\\ Arbitrary four-dimensional qubit rotation \cite{Li2018}\\ Non-Abelian geometric gates \cite{To2019}\end{tabular}        \\ \\ \\
Composite pulses                 & \begin{tabular}[c]{@{}l@{}}Pros: robust with high fidelity;\\ \ \ \ \ \ \ \ \ general and universal\\ Cons: fast increasing complexity with \\ \ \ \ \ \ \ \ \ extension of sequence lengths\end{tabular} & \begin{tabular}[c]{@{}l@{}}Potential applications:\\ Perform dynamically corrected gates \cite{XW2012,KK2012,JP2013} \\ Construct robust quantum gates \cite{CZ2017} \\ Measuring noise dynamics \cite{GT2013}\end{tabular}                                                       \\ \\ \\
Quantum optimal control \ \                 & \begin{tabular}[c]{@{}l@{}}Pros: precise and fast\\ Cons: only numerical solutions;\\ \ \ \ \ \ \ \  \ \ high control complexity \end{tabular}                                                        & \begin{tabular}[c]{@{}l@{}}Experimental advances:\\ Open-loop optimization in SiMOS qubits \cite{YC2019}\\ Close-loop optimization in GaAs-based \\ singlet-triplet qubits \cite{Ce2020}\end{tabular}                                      \\ \hline
\end{tabular}
\end{table*}

$\mathit{Composite \; pulses} $\quad CPs have been first developed and used in NMR \cite{CP1990}, and now is also applied in other areas including trapped ions, neutral atoms, QDs, nitrogen-vacancy centers in diamond, etc.\cite{Bo2022}. In semiconductor QD spin systems, CPs is a promising approach to realize robust and highly accurate coherent quantum control suppressing decoherence and cancelling errors.

Among semiconductor qubits, singlet-triplet qubits are particularly prominent due to their relatively long coherence times and potential to perform simple all-electrical control \cite{FK2017}. These qubits, where quantum information is encoded in the singlet-triplet spin subspace of two
electrons trapped in a double QD, are also insensitive to homogenous fluctuations of the magnetic field. However, there are two main sources of noises which can hardly be avoided in laboratory systems, hindering precise experimental manipulation. They are Overhauser noise arsing from the hyperfine-mediated spin-flip-flop processes, and charge noise stemming from fluctuations of environmental voltage \cite{DJ2008,DC2009}. Fortunately, these highly non-Markovian fluctuations are very slow compared to typical qubit operation times, thus they can be almost treated as  quasistatic noises.

It is challenging to directly apply traditional control techniques under a set of physical constraints in singlet-triplet qubits. However, one may perform dynamically corrected gates (DCGs) by using a special sequence of composite pulses,dramatically eliminating the leading order of errors or even addressing both relevant types of errors simultaneously, while fully respecting these experimental constraints \cite{XW2012,KK2012}. Moreover, Kestner J. P. et al. have extended the protocol to multi-qubit systems and demonstrated its potential of generating universal dynamically corrected operations on a large-scale quantum register \cite{JP2013}. In general, DCGs will also extend the coherence time to a much longer scale. And further research examined such models in a more realistic scenario with constraints of pulse generation \cite{XW2014}, e.g., in the presence of $1/f^\alpha$ noise.

Another solution to reduce the sensitivity of singlet-triplet qubits to charge noise is to use symmetric barrier control, which can maintain the symmetry between the dots and tune the exchange interaction by adjusting the electrostatic barrier \cite{FM2016}. It performs well in purified Si, but gives essentially no improvement for materials where nuclear spin noise dominates like GaAs. Zhang C. et al. proposed a new scheme that can implement gates up to ten times faster with a set of composite pulses, reducing the sensitivity to nuclear spin noise \cite{CZ2017}.

Such schemes may not apply to other semiconductor spin qubits. Hickman G. T.  et al. introduced a distinctive scheme for exchange-only qubit, which is experimentally implementable under realistic conditions within the unique physical constraints. \cite{GT2013}. Apart from cancelling noises, CPs can also be used to refocus exchange noise for measuring charge noise dynamics with exchange pulses \cite{En2015} [Fig. \ref{aplfig}(d)]. Fig. \ref{aplfig}(d) (middle) show the pulse sequence during control process. Different charge regions for the three dots are labeled by color and number. Singlet initialization (I) occurs during the transition from (1,0,1) to (2,0,1), which is followed by the evolution (E) to (1,1,1) and two charge-sensing measurement segments (M).

~\\

$\mathit{Quantum \; optimal \; control} $\quad QOC algorithms have already demonstrated their capabilities in many other physical platforms \cite{TU2016,NO2022,PJ2022,TM2011,JZ2018,Sv2014,AO2019,RW2017,FH2017}, yet there are limited applications in semiconductor qubits. There remain a wide range of potential uses of QOC methods in QD systems. Notably, Yang C.H. et al. experimentally showed that they could improve average singel-qubit Clifford gate error rates for silicon QD spin
qubits from the previous best results 0.$14$\% to $0.043$\% via pulse engineering \cite{YC2019} [Fig. \ref{aplfig}(e)]. They used GRAPE to identify pulses for qubit control that are robust against low-frequency detuning noise and this leads to a great improvement in decoherence time compared to traditional squire pulses. Furthermore, it is possible to characterize the noise by exploiting recent developments in randomized benchmarking \cite{JJ2018}. Fig. \ref{aplfig}(e) (middle) shows the optimal control sequences for the seven basic types 
of Clifford gate found through GRAPE iteration. Fig. \ref{aplfig}(e) (right) gives the randomised benchmarking results of both the original (square) pulses scheme $\mathbf{S} $, and the improved pulses scheme $\mathbf{O} $.The incoherence part indicates indicates the amount of infidelity (grey) that results from incoheronce noise.And the green lines are Sequential Monte Carlo estimates of the pulse fidelities, which can show the statistical distribution of fidelities.

Even if the models used for pulse optimization are not sufficiently accurate for the direct (``open-loop'') experimental application of the results, they can be a useful starting point for further fine tuning in a closed loop with
feedback from an actual experiment. Cerfontaine P. et al. achieved accurate control of GaAs-based singlet-triplet qubits encoded in two-electron spins with a fidelity of $99.50 \pm 0.04$\%, using careful pulse optimization and closed-loop tuning \cite{Ce2020}. They started with numerically optimized pulses, then the remaining inaccuracies in these optimized pulses can be removed by a closed-loop gate
set calibration protocol (GSC), which allows the iterative tune-up of gates using experimental feedback.

In addition to applying QOC algorithms in modeled or practical systems based on measurement, QOC also provides analytical methods for Pontryagin's maximum principle, geometric approaches, adiabatic evolution and dynamical decoupling, to name a few \cite{Ma2023}. As an example, quantum sensing has been performed with
optimal control theory (open-loop) pulses for improved magnetometry, with single-spin sensors \cite{Mu2018} . These sensors may be implemented in electrical readout of single spins on semiconductor QDs.

\section{Summary And Outlook}\label{outlook}
Quantum control plays a critical role in the development of quantum information processing, as many applications in quantum technology depend on the precise manipulation of quantum systems, e.g.,preparation of stable quantum gates, creation of high-fidelity robust entanglement. Since quantum applications depend strongly on the specific properties of the underlying quantum hardware, and semiconductor materials have led the way in industrial-scale engineering development, semiconductor QD spin qubits are particularly promising for integrated quantum computing. In this review, we introduced the basic concepts of quantum control problems, followed by a quick overview of the five widely-used quantum control methods. We also provided descriptions in terms of their capacities, advantages, disadvantages. Finally, we discussed the practical aspects of these methods and highlighted their demonstrated and potential applications in semiconductor spin qubits. A quick summary of these methods with comparison is shown in Table. \ref{comparison}.

Although over the past few years, much progress has been made in the design and demonstration of high-fidelity control over these qubits \cite{TO2016,FM2016,En2015,DK2015}, there still remain a lot of challenges in precisely controlling each type of qubit. For example, the challenges associated with controlling the local AC magnetic field for single spin rotations, as well as the need for inhomogeneous magnetic field control for singlet-triplet spin qubits, complicate their experimental implementation in QDs \cite{MP2008,JP2005}. The charge noise arising from environmental voltage fluctuation may lead to decoherence. With the use of advanced control methods, one may perform arbitrary quantum gate operation in a robust way suppressing decoherence and cancelling errors \cite{DC2009,NT2011}.

In practice, various complications need to be taken into account under realistic experimental conditions. For example, the exchange coupling among qubit states may depend on the detuning which can make a large impact on the effect of charge noise. Another complication stems from the experimental fact that control pulses cannot be modulated exactly as the same as what we designed. They usually contain a finite rise time between adjacent parts without a smooth transition. Nevertheless, one can start with the pulse parameters derived from assuming perfect pulses and then look for optimal values around them to obtain noise-resistant sequences. Furthermore, the noise cannot always be treated as static as assumed in many models, even though under a rather fast gate operation. To make the schemes experimentally feasible, it is also crucial to shorten the total gate operation times as much as possible, despite these issues that arise in real system.

The presence of noise from various experimental complications, however, should not diminish the applicability of these protocols. Numerous studies have shown that their control methods are robust against different types of errors. Therefore, we may expect that quantum control methods will play an increasingly important role in designing precise control pulses that remain robust against various noise under complex physical constraints \cite{BT2021,XW2014}. Additionally, measurement-based feedback control is a promising and novel avenue to perform ideal quantum control tasks, especially when considering the complexity of noises. 

\section{Acknowledgement}
We acknowledge support by the National Natural Science Foundation of China (No. 12174379, No. E31Q02BG), the Chinese Academy of Sciences (No. E0SEBB11, No. E27RBB11), the Innovation Program for Quantum Science and Technology (No. 2021ZD0302300) and Chinese Academy of Sciences Project for Young Scientists in Basic Research (YSBR-090).

\end{document}